\documentclass[12pt]{iopart}
\usepackage[dvips]{graphicx}
\usepackage{iopams}
\usepackage{setstack}
\newcommand{\id}{\hat{\mathbf{1}}_{N}}
\newcommand{\wek}[1]{{\vec{\bf {#1}}}}

\begin{document}

\title[Boundary states for $N \times N$ density matrices]
{Stratificational and antipodean properties of
boundary states for  $N \times N$ density matrices}
\author{S Kryszewski, M Zachcia{\l}, {\L} Szparaga}
\address{Institute of Theoretical Physics and
         Astrophysics, University of Gda{}\'{n}sk,
         ul. Wita Stwosza 57, 80-952 Gda{\'{n}}sk, Poland.}
\ead{fizsk@univ.gda.pl, dokmz@univ.gda.pl}
\begin{abstract}
We investigate the space of $N \times N$ dimensional density
matrices. We show that there exist strata such that boundary
states $\rho_{p}$ with $p$ zero eigenvalues lie on or
outside the spheres with radii $r_{p}=\sqrt{p/N(N-p)}$.
Moreover, we show that if in a certain direction there is
a boundary state with $q=N-p$ equal eigenvalues, then in the
opposite (antipodean) direction exists a boundary state with
$p=N-q$ equal eigenvalues.
\end{abstract}
\submitto{\JPA}


\section{Introduction}

The present renewed interest in the geometry of density matrices
describing $N$-level quantum systems is largely due to the
advances in the fundamentals of quantum mechanics and in
quantum information theory \cite{keyl}. The recent monograph
by Bengtsson and {\.Z}yczkowski \cite{bezy} gives an excellent
review of the subject and also very extensive bibliography.
Nevertheless, there are still many important open questions.
It is relatively easy to account for the required hermiticity
and normalization of density matrices. But the requirement
of positivity (strictly speaking: semi-positivity) causes
serious problems. One of the methods of investigation of
density matrices uses a concept of generalized Bloch
vectors. This is due to the fact that the space of density
matrices is in one-to-one correspondence with the space
of these vectors (see, for example \cite{kim,bkh,kryz,bert}).
The recent paper by
Kimura and Kossakowski \cite{kiko} proceeds along these
lines. These authors construct the spherical-coordinate
point of view in the space of Bloch vectors.
To our minds, perhaps the term {\em directional} would be better.
Kimura and Kossakowski derive inequalities which must be
satisfied by the lengths of Bloch vectors in the specific
directions. Moreover, they show that if there is a pure
state in certain direction then in the opposite one there
occur states with at most one zero eigenvalue.

The aim of this work is to simplify and generalize the results
of Kimura and Kossakowski. We stress that we deal only
with $N$-dimensional quantum systems described by
$N \times N$ density matrices.

In the next section we briefly present the basic theoretical
notions necessary in our analysis. We will formulate
the expressions giving the metric in the space of density
operators and hence the tool to measure distances which
can also be expressed in terms of lengths of Bloch vectors.
In the third section we will apply introduced ideas
and show that the space of density matrices is stratified
(that is possesses, colloquially speaking, onion-like structure).
The fourth section is devoted to a presentation of the directional
representation of density matrices. This is a brief review
of some results of Kimura and Kossakowski \cite{kiko} which
seem to be essential for the completeness of this work.
In the fifth section we discuss and generalize the antipodean
properties anticipated in \cite{kiko}. In the last section
we give some additional comments and remarks.

For further convenience, we present here a very simple lemma
which is due to Harriman \cite{har} and is quite useful
in further discussion.

{\bf Lemma.} Let $a_{j} \in \mathbb{R}, (j=1,2,\ldots,n $)
and $\sum_{j=1}^{n} a_{j}=1$ (note that numbers $a_{j}$ need
not be positive). Denoting $\sum_{j=1}^{n} a^{2}_{j} =A$,
one has:
\numparts {\label{b01}}
\begin{eqnarray}
   (i) &\quad& A \geq 1/n;
\label{b01a} \\
  (ii) &\quad& A = 1/n \qquad \Leftrightarrow \qquad a_{j}=1/n,
       \quad (j=1,2,\ldots,n).
\label{b01b} \end{eqnarray}
\endnumparts
{\bf Proof:}
Consider the sum $S$ defined as follows
\begin{equation}
  S = \sum_{j=1}^{n}
      \left( a_{j} - \frac{1}{n}\right)^{2}
    = \sum_{j=1}^{n}
      \left( a^{2}_{j} - \frac{2a_{j}}{n} + \frac{1}{n^{2}}\right)
    = A - \frac{1}{n},
\label{b02} \end{equation}
where we used the condition imposed upon the sum of numbers $a_{j}$.
Then, we have $A = S + 1/n$. Sum $S$ is a sum of squares so
it is nonnegative, then $A \geq 1/n$,
which proves \eref{b01a}. Let us now assume that $A=1/n$. Then
relation \eref{b02} entails $S=0$, which means
that all $a_{j} = 1/n$. Conversely, taking $a_{j}=1/n$ we see
that (\ref{b02}) gives $A=1/n$.
This completes the proof of the lemma.

\section{Theoretical framework}

Let us denote the maximally mixed state as
\begin{equation}
   \rho_{max}
   = \mathrm{diag}
     \left( \frac{1}{N},\frac{1}{N}, \ldots,\frac{1}{N} \right)
   = \frac{1}{N} \id,
\label{b06} \end{equation}
where $\id$ is a unit $N \times N$ matrix. Hilbert-Schmidt
distance between an arbitrary density matrix and $\rho_{max}$
is given as
\begin{equation}
   d^{2}(\rho,\rho_{max})
   = \mathrm{Tr} \left\{ \left( \rho - \rho_{max} \right)^{2}
     \right\}.
\label{b07a} \end{equation}
The trace is invariant with respect to unitary transformations
and so is $\rho_{max}$. Then we choose such a representation
in which $\rho = \mathrm{diag}(\lambda_{1},\lambda_{2},\ldots,%
\lambda_{N})$, where $\lambda_{j}$ are the eigenvalues of the
density matrix and satisfy the requirements:
$(i)~\lambda_{j} \in \mathbb{R}$ (hermiticity),
$(ii)~\sum_{j=1}^{N} \lambda_{j}= 1$ (normalization),
$(iii)~\lambda_{j} \geq 0$ (semi-positivity).
Using this representation we have
\begin{equation}
   d^{2}(\rho,\rho_{max})
   = \sum_{j=1}^{N} \left( \lambda_{j} - \frac{1}{N} \right)^{2}
   = \sum_{j=1}^{N} \lambda_{j}^{2} - \frac{1}{N}.
\label{b08} \end{equation}
Since $\mathrm{Tr} \{ \rho^{2} \} \leq 1$ we immediately get
\begin{equation}
      d^{2}(\rho,\rho_{max}) \leq \frac{N-1}{N}.
\label{b09} \end{equation}
This means that all density matrices lie on or within a sphere
(in the space of $N \times N$ hermitian and normalized matrices)
with the center at $\rho_{max}$ and radius
\begin{equation}
    R_{L}=\sqrt{\frac{N-1}{N}}.
\label{b09a} \end{equation}
We will call this sphere a large one. On the other hand,
it is well known that for $N \ge 3$, not all matrices within
this sphere are true density ones. The sphere also contains
hermitian and normalized
but nonpositive matrices. It is also known (see \cite{har})
that there exists
a small sphere (centered also at $\rho_{max}$) with a radius
\begin{equation}
    R_{S} = \sqrt{\frac{1}{N(N-1)}},
\label{b10} \end{equation}
inside which there are only density matrices. The nonpositive ones
can be found only outside a small sphere.

Above considerations were made by reference to the eigenvalues
of the density matrices. Similar ones can be done in terms of
the generalized Bloch vectors which are introduced as follows.

The basis in the space of hermitian  ${N\times N}$ matrices
consists of $N^{2}$ elements. We take the one consisting
of the unit matrix $\id$ and of $N^{2}-1$ traceless, hermitian
and orthonormal matrices $\left\{T_{j}\right\}^{N^{2}-1}_{j=1}$.
There are various possibilities of the choice of such a basis,
see the review \cite{bert}.
We assume that orthogonality holds in the Hilbert-Schmidt sense,
ie., ~$\mathrm{Tr} \left\{ T_{j}T_{k}\right\}=\delta_{jk}$.
Then, an arbitrary density operator for a $N$-level system can
be written as
\begin{equation}
    \rho
    = \frac{1}{N}\id + \sum_{i=1}^{N^{2}-1}V_{i} \, T_{i}
    = \frac{1}{N}\id + \wek{V} \cdot \wek{T},
\label{b13} \end{equation}
where $(V_{1},V_{2},\ldots,V_{N^{2}-1}) \in %
\mathbb{R}^{N^{2}-1}$  is called a generalized Bloch vector.
Representation \eref{b13} ensures normalization and hermiticity.
Requirement of positivity imposes restrictions upon vector
$\wek{V}$. To our knowledge, their general form is
not known although some particular cases \cite{kim,bkh,kryz}
were studied in much detail. Discussion of this point, however,
goes beyond the scope of the present work.

Bloch vector can be used to express the distance between an
arbitrary density matrix and the maximally mixed one
\begin{equation}
   d^{2}(\rho,\rho_{max})
   = \mathrm{Tr} \left\{ \left( \rho - \rho_{max} \right)^{2}
                 \right\}
   = \sum^{N^{2}-1}_{i,j=1} V_{i} V_{j}  \delta_{ij}
   = | \wek{V} |^{2}.
\label{b14} \end{equation}
as it follows from tracelessness and orthonormality of matrices
$T_{j}$. Comparing this relation with \eref{b08} and \eref{b09}
we can write
\begin{equation}
   | \wek{V} |^{2}
   = \sum_{j=1}^{N} \lambda_{j}^{2} - \frac{1}{N}
   \leq \frac{N-1}{N}.
\label{b15} \end{equation}
Therefore, the length of the Bloch vector is not arbitrary.
Bloch vectors $\wek{V} \in \mathbb{R}^{N^{2}-1}$
lie on or within a sphere with radius $R_{L}$. Note that
the space of Bloch vectors is different from the space of density
matrices, although relation \eref{b13} establishes a one-to-one
mapping between these spaces.

\section{Stratification of the Bloch space}

Boundary states are specified as such density matrices which
possess at least one eigenvalue equal to zero.
Let $\rho_{p}$ denote a boundary state with $p$ zero eigenvalues
and with $q=N-p$ nonzero positive ones: $\lambda_{1},\ldots,%
\lambda_{q}$ (obviously $\sum_{j=1}^{q} \lambda_{j} = 1$,
due to normalization). For such a boundary state relation
\eref{b08} gives
\begin{equation}
   d^{2}(\rho_{p},\rho_{max})
   = \sum_{j=1}^{q} \lambda_{j}^{2} - \frac{1}{N}.
\label{b20} \end{equation}
Due to lemma \eref{b01a} we know that $\sum_{j=1} %
\lambda_{j}^{2} \geq 1/q$.
Therefore
\begin{equation}
   d^{2}(\rho_{p},\rho_{max})
   \geq \frac{1}{q} - \frac{1}{N}
   =\frac{N-q}{Nq} = \frac{p}{N(N-p)}.
\label{b21} \end{equation}
This inequality allows us to draw several conclusions.
Some of them are obvious and well known, but we quote them
just for completeness of the reasoning. Some other
seem to be new.

Boundary states with just one zero eigenvalue ($p=1$) lie on
or outside the small sphere  (centered at $\rho_{max})$
of radius $R_{S}=\sqrt{1/N(N-1)}$. In other words,
the small sphere contains density matrices with all $N$
eigenvalues greater than zero. It is worth noting, that nonpositive
but hermitian and normalized matrices can also be found
just outside this sphere \cite{har}.

Pure states (with $\lambda_{1}=1$, all other eigenvalues are zeroes,
so that $p=N-1$) satisfy
\begin{equation}
   d^{2}(\rho_{pure},\rho_{max})
   \geq \frac{N-1}{N}.
\label{b23} \end{equation}
Obviously, this result together with relation \eref{b09}
imply that pure states must lie strictly on the surface
of a large sphere \cite{har}, so inequality \eref{b23} becomes an
equality. This explains why pure states are also extremal.
Pure states must lie on the surface of the large sphere
(which is a very well-known result) but the
converse is not true. There are (on the large sphere) matrices
which are not positive definite \cite{har}.

It seems to be a new conclusion that relation \eref{b21} implies
that the space of density matrices is stratified by concentric
spheres $S_{p}$ (centered at $\rho_{max}$) of radii
\begin{equation}
      r_{p}=\sqrt{\frac{p}{N(N-p)}},
      \qquad p=1,2,\ldots,N-1,
\label{b24} \end{equation}
and boundary states $\rho_{p}$ (with $p$ zero eigenvalues) lie
on or outside these spheres. We also note that for large and
small spheres, we have $R_{S} = r_{1}$ and $R_{L} = r_{N-1}$.

Lemma \eref{b01b} allows us also to say that inequality \eref{b21}
is minimized for the boundary state of the form
\begin{equation}
  \rho_{p}
  = \mathrm{diag}
    \biggl(
    \underbrace{\frac{1}{q},\frac{1}{q},\ldots,%
    \frac{1}{q}}_{q=N-p~\mbox{\footnotesize terms}},%
    \underbrace{0,\ldots,0\rule[-4mm]{0mm}{6mm} }
    _{p~\mbox{\footnotesize terms}}
    \biggr) \equiv R(q),
\label{b25} \end{equation}
so only boundary states $R(q=N-p)$ lie on the spheres $S_{p}$.
Other boundary states with $p$ zero eigenvalues lie outside these
spheres.

The discussed properties of the space of density matrices can
easily be translated into the language of Bloch vectors.
Firstly, density matrices with all nonzero eigenvalues correspond
to Bloch vectors with the length
\begin{equation}
  | \wek{V} |^{2} < \frac{1}{N(N-1)}.
\label{b26} \end{equation}
Secondly, boundary states with $p$ zero eigenvalues correspond
to Bloch vectors such that
\begin{equation}
    | \wek{V} |^{2} \geq \frac{p}{N(N-p)}.
\label{b27} \end{equation}
For boundary states $R(q)$ defined in Eq.\eref{b25} the above
inequality becomes an equality and we have
\begin{equation}
    | \wek{V}_{R(q)} |^{2}
    = d^{2}(R(q),\rho_{max})
    = \frac{N-q}{qN}
    = \frac{p}{N(N-p)}.
\label{b28} \end{equation}
All these lengths increase with growing number $p$.
Finally, pure states correspond to
$| \wek{V} |^{2} =(N-1)/N$ -- on the large sphere.

We have shown that the space of density matrices has stratified
structure. With the growth of the number of zero eigenvalues
the minimal distance of the given density matrix from the maximally
mixed state also increases.

\section{Directional representation}

In any vector space, a vector can be represented by its direction
and length. This is done writing
\begin{equation}
  \wek{V} = | \wek{V} | \wek{n},
\label{b31} \end{equation}
with directional vector $\wek{n}$ being normalized:
$\sum_{j=1}^{N^{2}-1} n_{j}^{2} = 1$. Then, instead of
re\-pre\-sen\-ta\-tion \eref{b13} we can take
\begin{equation}
    \rho
    = \frac{1}{N}\id + | \wek{V} | \wek{n} \cdot \wek{T}
    = \frac{1}{N}\id + | \wek{V} | T_{\wek{n}},
\label{b32} \end{equation}
where $T_{\wek{n}} = \sum_{j=1}^{N^{2}-1}n_{i} \, T_{i}$.
Representation \eref{b32} may be called a directional (or spherical)
one. It was introduced and investigated by Kimura and Kossakowski
\cite{kiko}. Let us note that matrix $T_{\wek{n}}$ is hermitian and
has the following properties
\begin{equation}
   \mathrm{Tr} \{ T_{\wek{n}} \} = 0,
   \qquad
   \mathrm{Tr} \{ T_{\wek{n}}^{2} \} = 1,
\label{b33} \end{equation}
which follow from the tracelessness and orthonormality of $T_{j}$
matrices. From the above relations it follows that matrix
$T_{\wek{n}}$ possesses positive and negative eigenvalues
which we will denote by $\{ \mu_{k} \}_{k=1}^{N}$.
They can be ordered in a natural way
\begin{equation}
   0 < \mu_{1} \ge \mu_{2} \ge \ldots\ldots \ge \mu_{N} < 0.
\label{b34} \end{equation}
Subsequently, we can take the diagonal representation of the
$T_{\wek{n}}$ matrix in Eq. \eref{b32}. The corresponding
density matrix becomes also diagonal and its matrix elements become
\begin{equation}
    \rho_{jj} = \frac{1}{N} + | \wek{V} | \mu_{j}
    \qquad j=1,2,\ldots,N,
\label{b35} \end{equation}
which must be nonnegative. The most restrictive relation is
obtained for the smallest (negative) eigenvalue
$\mu_{N} = - |\mu_{N}|$ (according to \eref{b34}). Inserting
this eigenvalue into relation \eref{b35} one finds that
\begin{equation}
    | \wek{V} | \leq \frac{1}{N|\mu_{N}|}.
\label{b37} \end{equation}
This estimate for the length of the generalized Bloch vector
constitutes one of the main results of the work by Kimura and
Kossakowski \cite{kiko}. This is obtained here in a manner which
seems to be simpler and more intuitive. Next, these authors
investigated the minimal and maximal possible values of $\mu_{k}$.
They have obtained two cases
\begin{equation} \fl
    \mu_{1} = \mu_{max} =\sqrt{\frac{N-1}{N}},
    \quad \mathrm{and} \quad
    \mu_{k} = - \frac{1}{\sqrt{N(N-1)}},
    \quad k = 2,3,\ldots,N.
\label{b38} \end{equation}
\begin{equation} \fl
    \mu_{k} = \frac{1}{\sqrt{N(N-1)}},
    \quad k = 1,2,\ldots,N-1
    \quad \mathrm{and} \quad
\mu_{N} = \mu_{min} = - \sqrt{\frac{N-1}{N}}.
\label{b39} \end{equation}
In the first case \eref{b38}, from inequality \eref{b37}we obtain
\begin{equation}
    | \wek{V} | \leq \sqrt{\frac{N-1}{N}},
\label{b40} \end{equation}
which is not unexpected. Next, from relations \eref{b35}
and \eref{b38} we get
\begin{eqnarray}
    \rho_{11}
    &=& \frac{1}{N} + | \wek{V} | \sqrt{\frac{N-1}{N}},
\nonumber \\
    \rho_{kk}
    &=& \frac{1}{N} - | \wek{V} | \sqrt{\frac{1}{N(N-1)}},
    \qquad (k=2,\ldots,N).
\label{b41} \end{eqnarray}
Allowing for maximum value of $| \wek{V} |$ as implied by
\eref{b40} we have
\begin{equation}
  \rho_{11} = 1, \qquad
  \rho_{kk} = 0, \qquad (k=2,\ldots,N),
\label{b42} \end{equation}
which corresponds to a pure state.

The second case described by eigenvalues \eref{b39} yields
\begin{equation}
    | \wek{V} | \leq \sqrt{\frac{1}{N(N-1)}},
\label{b43} \end{equation}
so it corresponds to density matrices on or within a small
sphere. Then, from relations \eref{b35} we find
\begin{eqnarray}
    \rho_{kk}
    &=& \frac{1}{N} + | \wek{V} | \sqrt{\frac{1}{N(N-1)}},
    \qquad (k=1,\ldots,N-1),
\nonumber \\
    \rho_{NN}
    &=& \frac{1}{N} - | \wek{V} | \sqrt{\frac{N-1}{N}}.
\label{b44} \end{eqnarray}
Taking maximum value of $| \wek{V} |$ from inequality \eref{b43}
we have now
\begin{equation}
  \rho_{kk} = \frac{1}{N-1}, \quad (k=1,\ldots,N-1),
  \qquad
  \rho_{NN} = 0.
\label{b45} \end{equation}
So we arrive at a boundary state with one zero eigenvalue
lying on the surface of the small sphere. This summarizes
the basic results of Kimura and Kossakowski \cite{kiko}.
On the other hand one may ask what interesting facts can be
inferred. The answer is in the antipodean properties.
This will be discussed in the next section.
But before doing this, let us make two additional remarks.

Firstly, we note that when inequality \eref{b37} becomes
an equality then the cor\-res\-pon\-ding density matrix \eref{b32}
is a boundary state. Indeed, in such a case, relation \eref{b35}
for $j=N$ gives $\rho_{NN} = 0$. Moreover, eigenvalue $\mu_{N}$
may be degenerate (see, for exam\-ple \eref{b38}).

Secondly, let us construct a directional matrix
$T_{\wek{n}}$ which describes the boundary state $R(q)$
(Eq.\eref{b25}). According to representation \eref{b32}
we now write
\begin{equation}
   R(q) = \frac{1}{N} \id  + | \wek{V}_{R(q)} | T_{\wek{n}}.
\label{b46} \end{equation}
Taking the length of Bloch vector as in \eref{b28} and by
definition \eref{b25} we get
\begin{equation} \fl
   T_{\wek{n}}
   = \mathrm{diag}
     \biggl( \underbrace{\sqrt{\frac{N-q}{qN}},
                  \ldots,\sqrt{\frac{N-q}{qN}}
     }_{q~\mbox{\footnotesize terms}},
     \underbrace{-\sqrt{\frac{q}{N(N-q)}},
          \ldots,-\sqrt{\frac{q}{N(N-q)}}
     }_{p=N-q~\mbox{\footnotesize terms}} \biggr),
\label{b47} \end{equation}
as a directional matrix corresponding to the boundary state
$R(q)$. Note that the first $q$ elements of this matrix
are equal to $|\wek{V}_{R(q)}|$ as given in \eref{b28}.

\section{Antipodean properties}

The antipodes are defined as follows. If a unit vector $\wek{n}$
defines certain direction, then vector $(-\wek{n})$ specifies
the opposite -- an antipodean direction. Therefore, if a density
matrix is given by its directional representation as
in \eref{b32}, then its antipodean counterpart is of the form
\begin{equation}
    \rho_{a}
    = \frac{1}{N}\id + | \wek{V} | T_{-\wek{n}}
    = \frac{1}{N}\id - | \wek{V} | T_{\wek{n}},
\label{b51} \end{equation}
because it seems obvious that $ T_{-\wek{n}} = - T_{\wek{n}}$.

Let us consider a state $\rho_{a}(q)$ which is, by definition,
antipodean to the boundary state $R(q)$. Thus, we can
write
\begin{equation}
    \rho_{a}(q)
    = \frac{1}{N}\id - | \wek{V}_{a} | T_{\wek{n}},
\label{b55} \end{equation}
where $T_{\wek{n}}$ is given in \eref{b47} and $| \wek{V}_{a} |$
is a suitably chosen length of the Bloch vector. Thus, matrix
$\rho_{a}(q)$ is of the form
\begin{equation} \fl
   \rho_{a}(q)
   = \mathrm{diag}
     \Biggl(
     \underbrace{\frac{1}{N} - | \wek{V}_{a} |
                 \sqrt{\frac{N-q}{qN}}, \ldots
                }_{q~\mbox{\footnotesize terms}},
     \underbrace{\frac{1}{N} + | \wek{V}_{a} |
                 \sqrt{\frac{q}{N(N-q)}},\ldots
                }_{p=N-q~\mbox{\footnotesize terms}}
     \Biggr).
\label{b56} \end{equation}
The requirement of positivity entails a restriction on the
allowed lengths of the Bloch vector. This implies
\begin{equation}
   | \wek{V}_{a} | \le \sqrt{\frac{q}{N(N-q)}}.
\label{b57} \end{equation}
Taking the maximal allowed value of  $\left| \wek{V}_{a}  \right|$,
from \eref{b56} we get
\begin{equation}
   R_{a}(q)
   = \mathrm{diag}
     \biggl( \underbrace{0, \ldots, 0\rule[-4mm]{0mm}{6mm} }
     _{q~\mbox{\footnotesize terms}},
     \underbrace{\frac{1}{N-q}, \ldots,\frac{1}{N-q}
     }_{p=N-q~\mbox{\footnotesize terms}} \biggr)
   =R(p).
\label{b58} \end{equation}
which is antipodean with respect to the boundary state $R(q)$.
We note that any matrix of the form of \eref{b56}, with
$\left| \wek{V}_{a}  \right|$ satisfying inequality \eref{b57}
is antipodean to $R(q)$, while $R_{a}(q)=R(p)$ given by
Eq.\eref{b58} is the one most distant from the center, that is
from the maximally mixed state $\rho_{max}$.

This is a generalization of the conclusions derived by
Kimura and Kossakowski \cite{kiko} from the sets \eref{b38}
and \eref{b39} of the eigenvalues of
directional matrices. These two sets (up to numbering) clearly
correspond to antipodal directions. In one direction there
is a pure state ($p=N-1$ or $q=1$) while in the opposite
one we find at most the state \eref{b45}
with $p=1$ or $q=N-1$.

\section{Final remarks}

We have shown that in the space of $N \times N$-dimensional
density matrices there are concentric spheres $S_{p}$ (centered
at the maximally mixed state) with increasing radii
$r_{p}=\sqrt{p/N(N-p)}$. These spheres characterize the
stratification of boundary states $\rho_{p}$ with $p$ zero
eigenvalues. States $\rho_{p}$ lie on the surface $S_{p}$ or
outside them. We have also shown, that only special boundary states
$R(q=N-p)$ defined in \eref{b25} are on $S_{p}$. When a density
matrix possesses $q=N-p$ nonzero and unequal eigenvalues
it must be outside the sphere $S_{p}$.
The states situated inside spheres $S_{p}$ have less than
$p$ zero eigenvalues, perhaps even none (all nonzero).
In particular, $p=1$ corresponds to a small sphere
which contains only those density matrices which have all nonzero
eigenvalues. Conversely, $p=N-1$ correspond to pure states
which lie only on the surface of the large sphere.
This shows that the space of density matrices has a complex
stratified structure.

The complexity of this structure is amplified by antipodean
properties. If we assume that in a certain direction $\wek{n}$
we have a boundary state $R(q)$, then in the antipode the
corresponding density matrix $\rho_{a}(q)$ is given by
expression \eref{b56}. As long as the Bloch vector $\wek{V}_{a}$
has the length satisfying the sharp inequality \eref{b57}
we may expect that $\rho_{a}(q)$ has all nonzero eigenvalues.
When \eref{b57} becomes an equality, the antipodal matrix
$\rho_{a}(q)$ attains the form of a boundary state
$R(p=N-q)$ as given by Eq.\eref{b58}. So the point $R(q=N-p)$
on the surface $S_{p}$ corresponds to the antipodean point
$R(p=N-q)$ on the surface $S_{q}$.

As yet, we are not fully aware what are potential consequences
of the presented stratification of the space of $N \times N$
dimensional density matrices. Nevertheless, we feel that our
findings are interesting and deserving further investigations.


\ack
Partial support by Polish Ministry of Science and Education
PZB-MIN-008/P03/2003 is gratefully acknowledged.


\end{document}